\documentclass[12pt]{article}
\pdfoutput=1
\usepackage[table]{xcolor}
\usepackage{cite}
\usepackage{avant}
\usepackage{amssymb}
\usepackage{multirow,booktabs}
\usepackage{caption}
\usepackage{subcaption}
\usepackage{pifont}\usepackage{amsmath}
\usepackage[mathscr]{eucal}
\usepackage[all]{xy}
\usepackage{listing}
\usepackage{pdfpages}
\usepackage{graphicx}
\xyoption{arc}
\xyoption{curve}
\numberwithin{equation}{section}
		\newfont{\spfnt}{punk12}
		\newcommand\email[4]{#1@#2.#3.#4}

		\renewcommand\hat{\widehat}

		\newfont\sheafnt{rsfs10}

\begin{document}
			\title{Holographic charge density wave from D2-D8}
			\author{
				Nishal Rai \thanks{\email{nishalrai10}{gmail}{com}{}}
				and Subir Mukhopadhyay \thanks{\email{subirkm}{gmail}{com}{}} 					
			}
			\date{$^*$ {\small Department of Physics, SRM University Sikkim, 
5th Mile, Tadong, Gangtok 737102}\\
			$^\dagger$ {\small Department of Physics,Sikkim University, 6th Mile, Gangtok 737102} }
			\maketitle
			\begin{abstract}
				\small
				\noindent
				We have considered D2-D8 model and obtain a numerical solution that exhibits spatially modulated phases corresponding to a charge density wave and a spin density wave. We have analysed behavior of the free energy density for different values of the chemical potential and the magnetic field.  
			\end{abstract}
			\thispagestyle{empty}
			\clearpage

		\section{Introduction}

There are varieties of condensed matter systems that exhibit spontaneous breaking of spatial symmetry. In particular, some systems admit spatially modulated phases that correspond to the breaking of the translation symmetry and symmetry of the lattice in the ground states. Common examples of such phases are charge and spin density waves \cite{23,24}. Charge density waves were first predicted \cite{Pierls} for a weakly coupled one-dimensional system. Various metals admit spatially modulated charge density phases. They appear in strongly correlated systems as well and sometimes they are accompanied by spin density waves, as found in the striped phases of high T$_c$ cuprate superconductors \cite{vojta}. They are characterized by competing orders, which is believed to play an important role in the rich phase structure of high T$_c$ superconductors.  Such breaking of translation symmetry modifies the transport behaviors, {\it e.g.}  showing unconventional behaviors of DC and AC conductivities. 

Gauge/gravity duality offers a novel method to study the strongly correlated systems by translating problems in a strongly coupled d-dimensional field theory into phenomena of a weakly coupled gravity theory living in (d+1) dimension \cite{Maldacena:1997re,Gubser:1998bc,Witten:1998qj,review}. A number of works have appeared, where this duality has been applied for studying systems with broken symmetry. Solutions with explicitly broken symmetry can be obtained by introducing sources that depend on space, such as chemical potential. On the other hand, sometimes spatially homogeneous solutions in presence of large charge density or magnetic field develop instability leading to spatially modulated configurations. Systems with such instabilities arise in the context of holographic QCD as well as probe brane models. One of the mechanisms of such instability occurs in Maxwell-Chern-Simons theory  at non-zero momentum once an electric field is turned on \cite{1}. When coupled to gravity, RN-AdS black hole develops a similar instability in presence of an electric field, which depends on the non-zero momenta suggesting a phase transition to an inhomogeneous solution and thus signals spatially modulated solution in the dual theory \cite{2}.

Instabilities leading to spatially inhomogeneous solution have been studied in quite a number of different models in bottom-up approach \cite{4,5,12,13,3,6,7,16,17,14,31,15,Withers:2013kva,18,8,9}. Electrically charged AdS black hole with a neutral pseudoscalar was studied in four dimension\cite{4} and  with $SU(2)\times U(1)$ gauge theory was studied in five dimensions in \cite{5}. In particular, analysis of linearized perturbations shows\cite{5} that below a critical temperature, the black hole is unstable towards decay into a phase with non-zero momentum, implying a spatially modulated phase. A full-fledged spatially modulated solution of it was obtained numerically in \cite{12}.  Subsequently,  a black hole solution with spatially modulated  horizon was obtained numerically in Einstein-Maxwell-pseudoscalar model with backreaction in \cite{13}. On the other hand in magnetically charged black branes, with geometry AdS2 or AdS3, space-dependent instabilities were found in \cite{3}. Similar momentum dependent instabilities were shown in a region of parameter space \cite{6} for a magnetically charged AdS2 solution, appearing in Einstein-Maxwell-dilaton model with  hyperscaling violating geometry. \cite{7} considered a geometry interpolating between AdS2 and AdS4, having anisotropic Lifshitz scaling and hyperscaling violation in the intermediate with an electric field. Linearised analysis of instabilities of this model predicts that its endpoint may lead to a spatially modulated phase. This was further extended to two U(1) gauge fields \cite{16} and found to admit a pair density wave, which involves the intertwined order of a superconducting phase and charge density wave. Furthermore, Dirac equation was analyzed in this background \cite{17} to obtain spectral density and  Fermi surface and band gap were explored.
Einstein-Maxwell-dilaton model was also considered in \cite{14}, leading to a spatially inhomogeneous solution as a thermodynamically preferred phase. A similar axionic system was found to exhibit instability towards the formation of inhomogeneous stripes in the bulk and on the boundary  \cite{31}. Such solutions were further generalized to a checkerboard solution, which breaks translation symmetry spontaneously in two directions \cite{15}. 
A higher derivative gravity model along with a complex scalar and gauge field was considered in \cite{18}, giving rise to a solution with spontaneously breaking translation symmetry, which corresponds to a charge density wave at T=0 .

Spatially inhomogeneous models are analysed in the top-down approach as well.  In \cite{28} a D7 brane probe in the background of D3 branes with a black hole embedding was studied using linearised perturbation and it was found that above a critical density the system develops instability, which is similar to a charge density and a spin density waves. A holographic model based on a D8 brane probe in the background of the D2 brane was considered in \cite{Jokela:2011eb} and its conductivity and phase structure was analysed.  Subsequently, \cite{29} analysed fluctuations in this model and found instability leading to an inhomogeneous striped phase. At high enough charge density, the instabilities developed in homogeneous state of the D3-D7 model leads to a spatially modulated configuration, which is a CDW and an SDW \cite{10}. The phase space of the parameters and nature of the transition was also studied. Analysis of AC and DC conductivities of this model appeared in \cite{11} and it was found that by applying an electric field, the stripes slide. It was also shown that for large chemical potential, the ground state will be a combination of a CDW and an SDW. Field theory with spontaneous stripe order was considered in holographic setup \cite{11} and DC and AC conductivities with a small applied electric field were studied.
It was shown in \cite{38} that spontaneously broken spatially modulated ground state with explicit symmetry breaking in the form of ionic lattice etc. gives rise to pinning. Direct and optical conductivities were analysed in this model.
 
As mentioned above, a charge density wave along with  a spin density wave occurs in the striped phases of high T$_c$ cuprate superconductors \cite{vojta}. The aim of the present work is to study a probe D2-D8 brane model in a top-down approach, which admits both charge density and spin density waves. This D2-D8 brane model was introduced in \cite{Jokela:2011eb} and subsequently, an analysis of the linearised fluctuations of this model showed \cite{29} that once the charge density of the homogeneous black hole phase is greater than some critical value, it becomes unstable and leads to spatially modulated configuration. We have obtained a spatially modulated numerical solution for this model, and from the analysis of the free energy, we find that this solution becomes a thermodynamically preferred phase when chemical potential crosses a critical value. In this spatially modulated solution, the bulk fields are coupled with one another and from the study of the asymptotic behavior of the gauge field and a certain scalar field we find that it is a mixture of both a charge density wave and a spin density wave. The Nature of the phase transition turns out to be second order in absence of the magnetic field. We have studied the behavior of the free energy in the presence of the magnetic field as well and find that free energy increases with the magnetic field. For the non-zero magnetic field, the phase transition becomes first order and the critical value of the chemical potential increases. For a large value of the chemical potential, however, it encounters an electrical instability in the case of both zero and non-zero magnetic field.

The article is structured as follows. In the next section we review the D2-D8 model, discuss the equations and the numerical method employed to solve it. The results are given in section 3 and we conclude with a discussion in section 4.. 

			\section{D2-D8 model }
We consider a probe D8 brane embedded in the background of N D2 branes \cite{Jokela:2011eb,29}. In the limit of large $N$ and large `t Hooft coupling $g_sN >> 1$, the holographic field theory dual is super Yang-Mills (SYM) in $(2+1)$ dimensions, along with charged fermions following from low energy degrees of freedom of bi-fundamental strings and it can serve as a dual model for applications in condensed matter.

 The ten-dimensional near-horizon metric for a D2 brane is given by \cite{Jokela:2011eb,29}
  \begin{equation}
  ds^2 = L_0^2 [ r^{5/2} (-f(r) dt^2 + dx^2 + dy^2) + r^{-5/2} (\frac{dr^2}{f(r)} + r^2 dS_6^2)],
  \end{equation}
  where $L_0^5 = 6 \pi^2 g_s N l_s^5$ and $f(r) = 1 - (\frac{r_T}{r})^5$. The metric on the internal sphere $S^6$ is written as
  \begin{equation}
  dS_6^2 = d\psi^2 + \sin^2\psi (d\theta_1^2 + \sin^2\theta_1 d\phi_1^2 ) + \cos^2\psi (d\xi^2 + \sin^2\xi d\theta_2^2 + \sin^2\xi \sin^2\theta_2 d\phi_2^2).
  \end{equation} 
  The ranges of the angular coordinates are given by $0 \leq \psi \leq \pi/2$, $0 \leq \xi,  \theta_1, \theta_2 \leq \pi$, $0 \leq \phi_1 , \phi_2 \leq 2\pi$. The dilaton is given by $e^\phi = g_s (\frac{r}{L_0})^{-5/4}$. The 5-form potential is given by $C^{(5)} = c(\psi) L_0^5 dS_2 \wedge dS_3$, where $dS_2$ and $dS_3$ are volume form on the $S^2$ and $S^3$ respectively. The coefficient $c(\psi)$ depends on the angular variable and is given as $c(\psi) = \frac{5}{8} (\sin \psi - \frac{1}{6} \sin (3\psi) - \frac{1}{10} \sin{5\psi} )$. In the following we set $L_0=1$.
  
We will consider the D8 brane to fill up $t$, $x$ and $y$ directions, wrapped $S^2$ and $S^3$ and extended in $r$ directions. The embedding then is given by $\psi$ as a function of the radial coordinate $r$ and $x$. In order to stabilize the embedding, we have turned on a magnetic field on $S^2$ given by $2 \pi \alpha^\prime F_{\theta_1\phi_1} = L_0^2 b \sin\theta_1$  \cite{Jokela:2011eb,29}. In addition to that we have turned on a gauge potential $a_\mu$, along the worldvolume as well as along the radial direction $a_{t,x,y,r}(r,x)$, which in general depends on the world volume directions as well as the radial direction.  We have also turned on a constant magnetic field $h$ along the $xy$ direction, $2 \pi \alpha^\prime  F_{xy} = h$. In what follows, we will choose a radial gauge and work with $a_r=0$.

We will replace the radial variable $r$ by a compact variable $u$, which ranges between $0$ to $1$ and will scale $x$ and $a_\mu$ to avoid explicit dependence of $r_T$  as follows.
\begin{equation}
r=\frac{r_T}{u},\quad a_\mu = r_T \hat{a}_\mu,\quad x_\mu = \hat{x}_\mu r_T^{-3/2},\quad \hat{b}= b \sqrt{r_T}.
\end{equation}

The action consists of a Dirac-Born-Infeld term and a Chern-Simons term, which are given in terms of the rescaled variables, as follows.
\begin{equation}\begin{split}
S &= S_{DBI} + S_{CS},\\
S_{DBI} &= - T_8 \int d^9x ~ e^{-\phi} \sqrt{-det (g_{\mu\nu} + 2 \pi \alpha^\prime F_{\mu\nu})}\\
  & = - N r_T^2 \int du~ d\hat{x}~ \frac{1}{u^2}~\sqrt{A (A_1 + A_2 + A_3)},
\end{split}\end{equation}
where $N= 8\pi^3 T_8 V_{1,1}$, $V_{1,1}$ being the volume of spacetime in $t$ and $y$ direction. $A$, $A_1$,$A_2$ and $A_3$ are defined as follows:
\begin{equation}\begin{split}
A&= \cos ^6{\psi} \left(\sin ^4{\psi}+\frac{{\hat b}^2}{u}\right),\\
A_1 & = \frac{1}{u^5}[ 1 + u^2 f \psi_u^2 -  u^4 {\hat a}_{0u}^2  + u^4 f {\hat a}_{{\hat y}u}^2  + f u^4 {\hat a}_{{\hat x}u}^2] \\
A_2 & = \frac{\psi_{\hat x}^2}{u^2} - \frac{{\hat a}_{0{\hat x}}^2}{f} + {\hat a}_{{\hat y}{\hat x}}^2\\
A_3 &= - u^2 {\hat a}_{0u}^2 \psi_{\hat x}^2     - u^2  {\hat a}_{0{\hat x}}^2 \psi_u^2  + u^2 f {\hat a}_{{\hat y}u}^2 \psi_{\hat x}^2   -  u^4 {\hat a}_{0{\hat x}}^2  {\hat a}_{{\hat y}u}^2   +  u^2 f {\hat a}_{{\hat y}{\hat x}}^2 \psi_u^2   -   u^4 {\hat a}_{0u}^2 {\hat a}_{{\hat y}{\hat x}}^2 \\
& + 2 u^2 {\hat a}_{0u} {\hat a}_{0{\hat x}} \psi_u \psi_{\hat x} + 2 u^4 {\hat a}_{0u} {\hat a}_{0{\hat x}} {\hat a}_{{\hat y}u} {\hat a}_{{\hat y}{\hat x}} - 2 u^2 f {\hat a}_{{\hat y}u} {\hat a}_{{\hat y}{\hat x}} \psi_u \psi_{\hat x}.   
\end{split}\end{equation}
In order to keep expressions compact we have used $\psi_u = \frac{\partial \psi}{\partial u}$, ${\hat a}_{0{\hat x}} = \frac{\partial {\hat a_0}}{\partial {\hat x}}$, etc.
As can be seen through a straightforward analysis, ${\hat a}_{\hat x}$ appears through a term $(\partial_u {\hat a}_{\hat x})^2/f $ in $A_1$, does not mix with other fields and thus decouples from the rest of the system. So we have dropped ${\hat a}_{\hat x}$.

The Chern-Simons term is given by
\begin{equation}\begin{split}
S_{CS} &= - \frac{T_8}{2} (2 \pi \alpha^\prime)^2 \int C^{(5)} \wedge F \wedge F ,\\
&= - N r_T^2 \int du ~ d{\hat x} ~ c(\psi) ({\hat a}_{0u} {\hat a}_{{\hat y}{\hat x}} - {\hat a}_{0{\hat x}} {\hat a}_{{\hat y}u}).
\end{split}\end{equation}
Therefore, the full action can be written as
\begin{equation}\label{action}
S_{DBI} + S_{CS} = - N r_T^2 \int du d{\hat x}~ \frac{1}{u^2} [ \sqrt{A (A_1 + A_2 + A_3)} + u^2 c(\psi) ({\hat a}_{0u} {\hat a}_{{\hat y}{\hat x}} - {\hat a}_{0{\hat x}} {\hat a}_{{\hat y}u}) ] .
\end{equation}
The equations of motion are obtained from this action by taking variation of the fields, $\psi({\hat x},u)$, ${\hat a}_0({\hat x},u)$ and ${\hat a}_y({\hat x},u)$. They give rise to nonlinear partial differential equations in ${\hat x}$ and $u$ and are difficult to solve. Since the equations are quite long we have not given the explicit expressions. The action is invariant under ${\hat x} \rightarrow -{\hat x}$ and the solutions also share this symmetry. Furthermore, in absence of magnetic field there is a reflection symmetry $\hat{x}\rightarrow \frac{L}{2} - \hat{x}$. 

The boundary conditions follow from physical considerations. At the ultraviolet limit, given by $u=0$ the boundary conditions are given by
\begin{equation}\begin{split}\label{boundaryuv}
\psi({\hat x},0) &= \psi_\infty ,\quad\quad
\partial_u \psi({\hat x},0) = m_\psi ,\\
{\hat a}_0({\hat x},0) &= \mu, \quad\quad\quad
{\hat a}_y(x,0) = h {\hat x},
\end{split}
\end{equation}
where $\mu$ is the chemical potential and we have turned on a constant magnetic field $h$ in the $xy$-plane of the world volume.  $\psi_\infty$ is the asymptotic value of the field $\psi$ at $u\rightarrow 0$, which we have chosen to be constant and $m_\psi$ represents mass of the fermion.

At the infrared limit, $u\rightarrow 1$ the boundary conditions follow from the regularity of the solutions. Since $A_2$ contains a term $\frac{(\partial_{\hat x} {\hat a}_0)^2}{f}$ and since $f$ vanishes at horizon, $\partial_{\hat x} {\hat a}_0$ has to vanish over there, implying ${\hat a}_0$ is a constant. Furthermore, the constant needs to vanish so that ${\hat a}$ is a well-defined one-form at the horizon. Therefore, we impose 
\begin{equation}\label{boundaryir}
{\hat a}_0({\hat x},1)=0.
\end{equation}
In addition, we have expanded the equations of motion corresponding to $\psi$, $a_0$ and $a_y$ around the horizon $u=1$ and consider the leading order equations in the expansions.

Since we are looking for a spatially modulated solution, we impose periodic boundary condition along ${\hat x}$ direction for all the fields:
\begin{equation}
\psi({\hat x} +L) = \psi({\hat x}),\quad\quad
{\hat a}_0({\hat x}+L) = {\hat a}_0({\hat x}), \quad\quad
{\hat a}_y({\hat x}+L) = {\hat a}_y({\hat x}).
\end{equation}

One can obtain the asymptotic behavior of the various fields considering $u\rightarrow 0$ limit. From the equations of motion, it turns out that the asymptotic behavior of $a_0$ and $\psi$ can be given by,
\begin{equation}
a_0(\hat{x},u) = \mu + d({\hat x}) u^2 + ..., \quad\quad 
\psi(\hat{x},u) = \psi_\infty + m_\psi u - c_\psi({\hat x}) u^3 + ....\label{asymbehave}
\end{equation}
In the expansion of $a_0$, apart from the leading term which is given by the chemical potential $\mu$, we have identified coefficient in the subleading term $d({\hat x})$ with the charge density function in the boundary field theory. Since we have considered a spatially modulated solution, this has a nontrivial dependence on ${\hat x}$ and we define average of the charge density over the period by
\begin{equation}
<d> = \frac{1}{L} \int\limits_0^L d({\hat x}) ~d{\hat x}.
\end{equation}
 The amplitude of the charge density wave can be obtained by considering Max$(d({\hat x}) - <d>)$. 
Similarly, we have identified \cite{Jokela:2011eb,29} the coefficient of $u^3$ in the asymptotic expansion of the field $\psi$ with the fermion bilinear in the dual field theory and Max($c_\psi({\hat x})-<c_\psi>$) will play a similar role in the context of spin density wave in the boundary field theory. Average of the spin density is also defined likewise in terms of $c_\psi$.

Since the system of partial differential equations following from the action is quite difficult to solve we have used pseudospectral method. We have expanded each function along $u$ direction in terms of Chebyshev polynomial for the region $0\leq u \leq 1$. In the $x$ direction we have expanded each function in terms of Fourier series. The expansions of the different fields are given by
\begin{equation}
\begin{split}
\psi(\hat{x}, u) = \sum\limits_{j=0}^{N_u-1} \sum\limits_{k=0}^{N_x-1} \psi[j,k] T_j(2u-1) \cos \frac{2\pi k \hat{x}}{L} , \\
{\hat a}_0(\hat{x}, u) = \sum\limits_{j=0}^{N_u-1} \sum\limits_{k=0}^{N_x-1} a_0[j,k] T_j(2u-1) \cos \frac{2\pi k \hat{x}}{L} , \\
{\hat a}_y(\hat{x}, u) = \sum\limits_{j=0}^{N_u-1} \sum\limits_{k=0}^{N_x-1} a_y[j,k] T_j(2u-1) \sin \frac{2\pi k \hat{x}}{L} . \label{expansion}
\end{split}\end{equation}

We have chosen the grid of collocation points as consisting of $N_x + 1$ points along $\hat{x}$ direction from $\hat{x}=0$ to $\hat{x}= {L}$, evenly distributed over the range and $N_u+1$ points from $u=0$ to $u=1$ forming the Gauss-Lobatto grid. Substituting the expansions in the equations of motion and boundary conditions and evaluating them at the collocation points reduces the system of partial differential equations to a set of algebraic equations of the coefficients in the expansion (\ref{expansion}). For numerical computation we have chosen $N_x=9$ and $N_u=11$. 

We have incorporated the first and third equations of boundary conditions at UV (\ref{boundaryuv}) in the choice of the coefficients and thus reducing the number of variables.  The rest of the boundary conditions are evaluated at the collocation points at $u=0$ and $u=1$ accordingly and the three equations of motion following from the action are evaluated at all the collocation points from the grid. We have chosen the number of collocation points in such a way so as to have the number of variables equal to the number of algebraic equations and solve them numerically.

For the numerical solution of the algebraic equations we have used the Newton-Rhapson method.  This set of algebraic equations has an additional difficulty stemming from the fact that it involves square root, due to the DBI part of the action. For a generic choice of initial value, it leads to complex solution. One has to choose the initial value judiciously so as to get real solutions for the coefficients. 

\section{Spatially modulated solution}

In this section, we will consider spatially inhomogeneous solutions of the equations of motion following from the action of the D2-D8 model (\ref{action}), which consists of three partial differential equations and four boundary conditions.  Furthermore, they being coupled partial differential equations, it is difficult to obtain an analytical solution. We have used the pseudospectral method as elaborated in the last section and obtain solutions to these partial differential equations numerically. Regarding the various parameters in the boundary conditions, we have made the following choices. In accordance with the analysis in \cite{Jokela:2011eb,29}, we have set the value of the field $\psi$ at the boundary $u=0$ to be $\psi_\infty =0$. The parameter $m_\psi$, that appears in the second of the equation (\ref{boundaryuv}) represents the mass of the fermions. 
As explained in \cite{Jokela:2011eb},  when $\psi(u)$ vanishes at a finite $u$ with finite $\psi^\prime$ D8 brane self-intersects, leading to conical singularity. For such a configuration, the massive modes which are not present in the DBI action may become tachyonic, leading to the breakdown of the present low energy approximation. Therefore, to ensure that the present model represents the D2-D8 model,  $\psi$ needs to be positive over the range of u and x.  In order to avoid such a breakdown, we have chosen $m_\psi$ to be positive and set its value to be $m_\psi=0.5$.  We choose $b=1$ and consider appropriate variations in $\mu$, $L$ and $h$ as mentioned in the following.
We begin with the condition that the magnetic field is absent and set $h=0$. Using the numerical procedure, we look for a spatially inhomogeneous solution by varying $\mu$ and $L$. It turns out that there are a number of branches of solutions and we have considered a specific branch of solution.

\begin{figure}[h]
			\centering
			\begin{subfigure}{8cm}
				\centering
				\includegraphics[width=7cm]{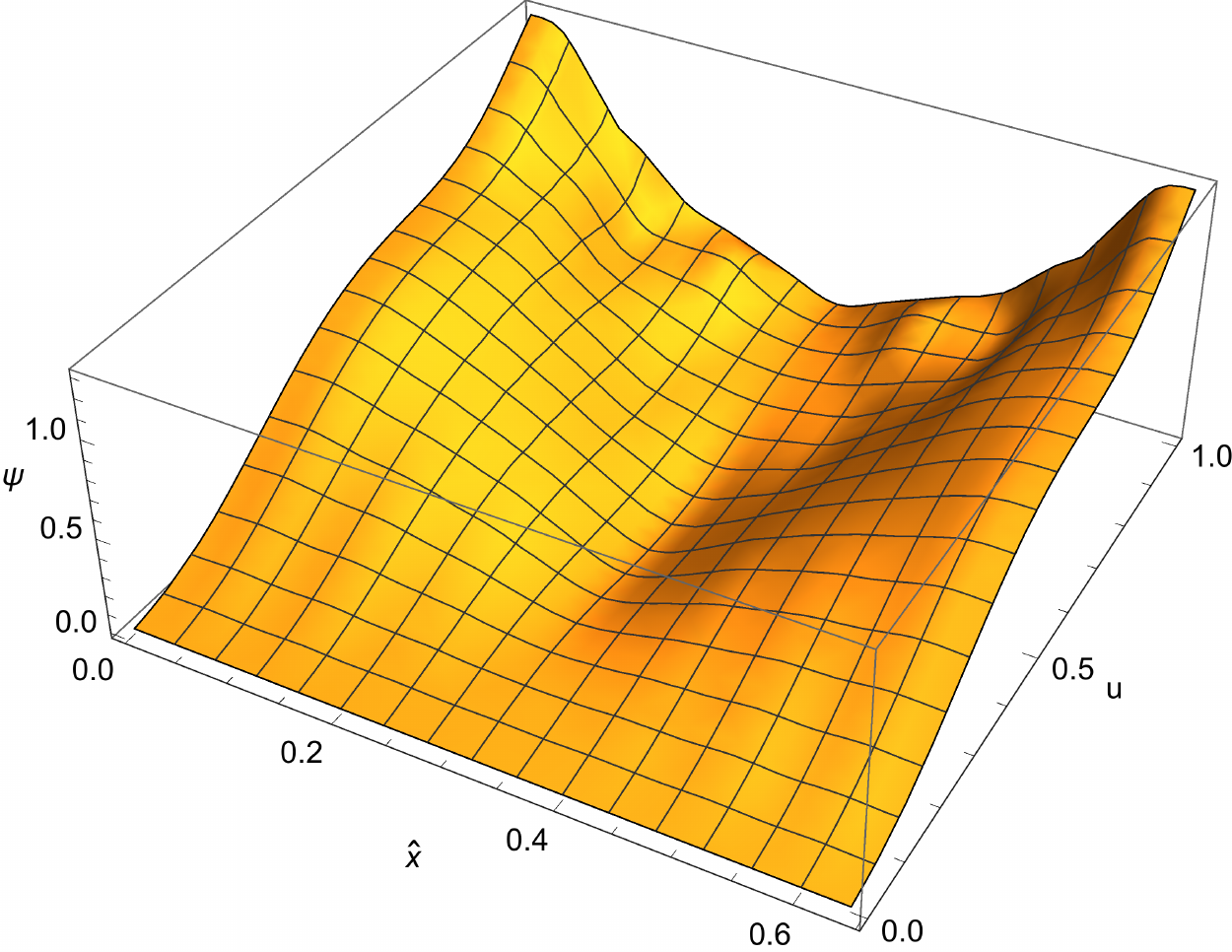}
						\caption{}
				\label{fig:1a}
			\end{subfigure}%
			\begin{subfigure}{8cm}
				\centering
				\includegraphics[width=7cm]{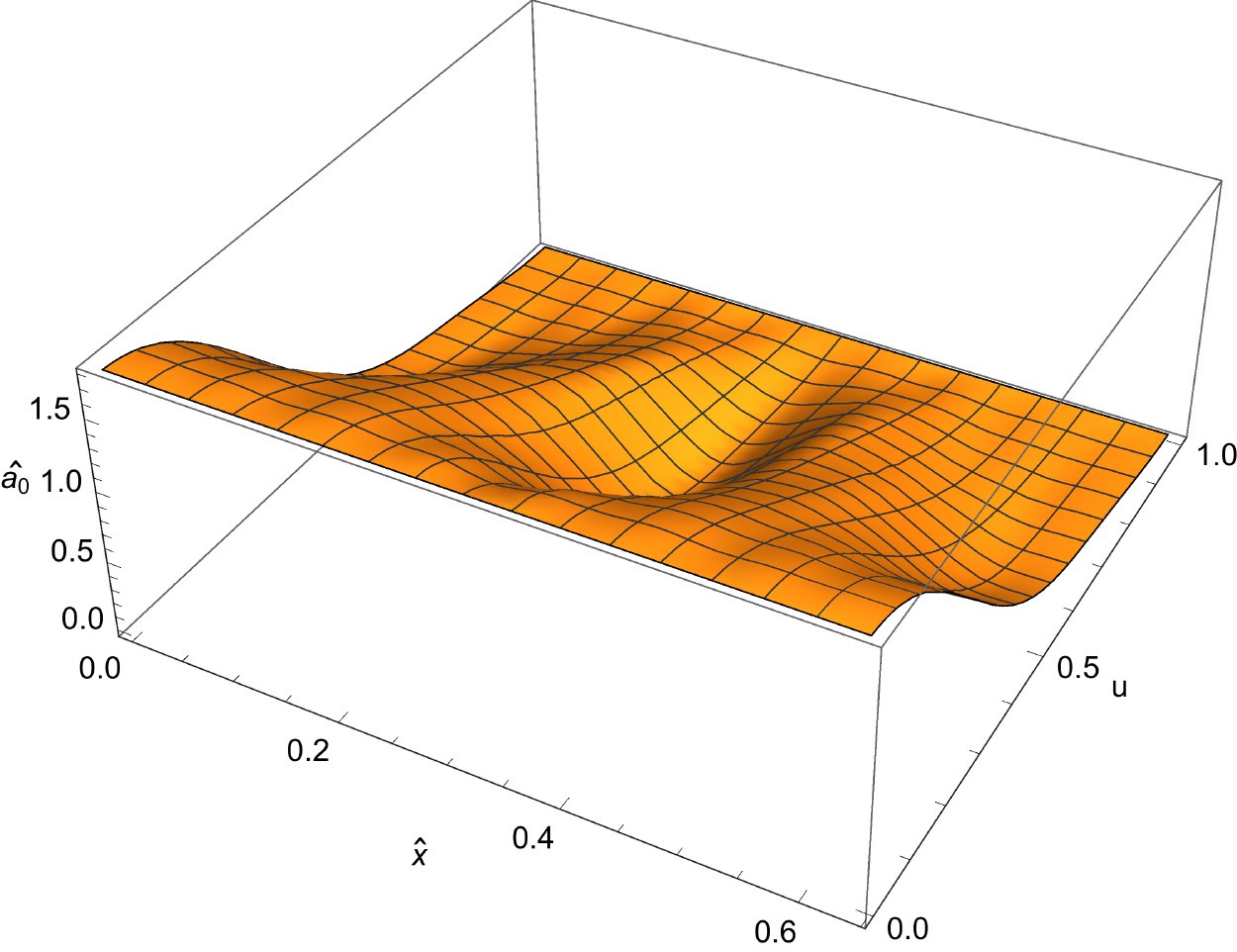}
						\caption{}
				\label{fig:1b}
			\end{subfigure}
			\caption{Plot of $\psi$ and $a_0$ vs. $\hat{x}$ and $u$ }
			\label{fig:plot1}
		\end{figure}

\begin{figure}[h]
			\centering
			\begin{subfigure}{8cm}
				\centering
				\includegraphics[width=7cm]{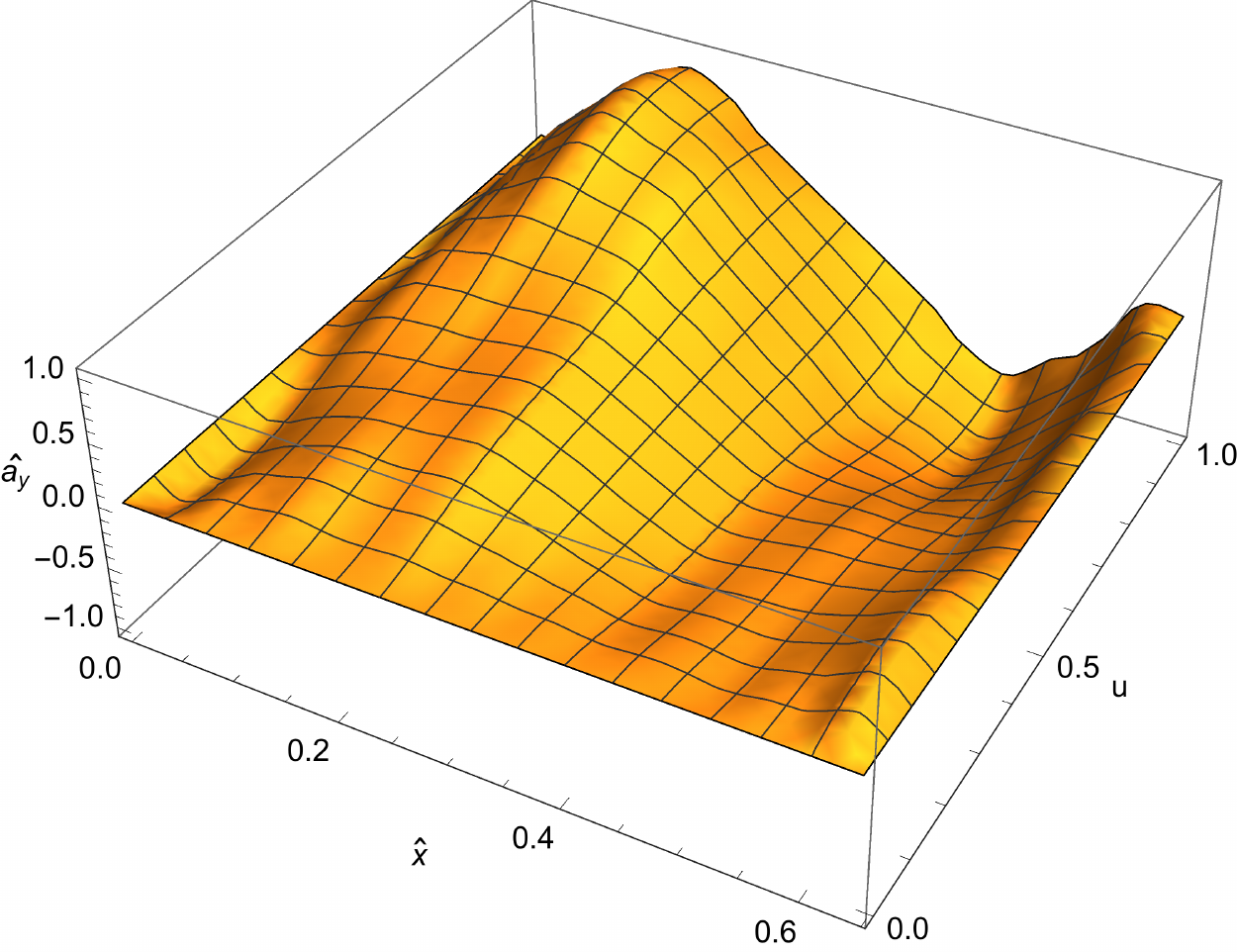}
						\caption{}
				\label{fig:2a}
			\end{subfigure}%
			\begin{subfigure}{8cm}
				\centering
				\includegraphics[width=7cm]{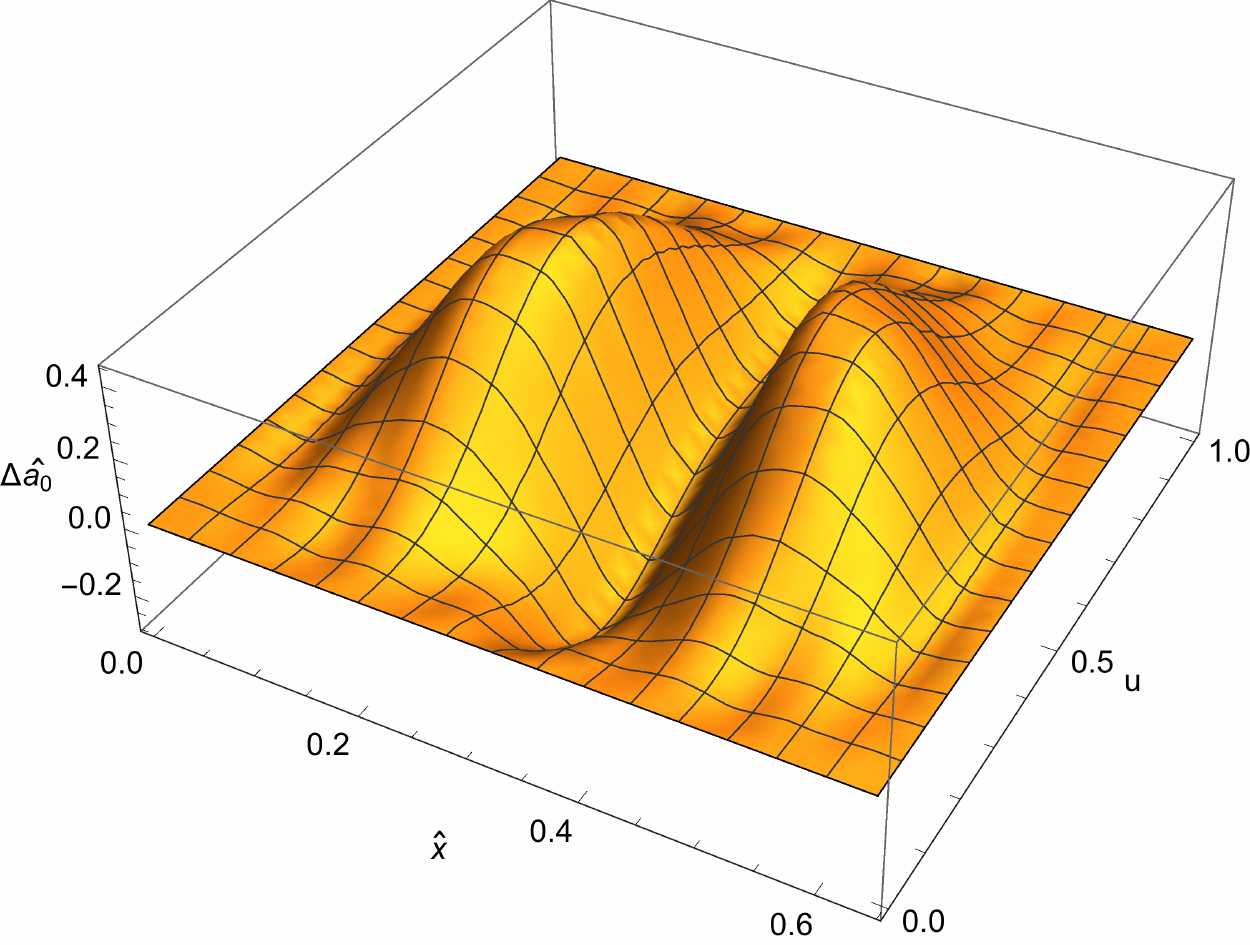}
						\caption{}
				\label{fig:2b}
			\end{subfigure}
			\caption{Plot of $a_y$ and the modulation in $\triangle a_0 = a_0(\hat{x},u) - a_0(0,u)$ vs. $\hat{x}$ and $u$}
			\label{fig:plot2}
		\end{figure}

A representative solution for $\psi$, ${\hat a}_0$ and ${\hat a}_y$ are given in figure \ref{fig:1a}, figure \ref{fig:1b}, and figure \ref{fig:2a} respectively. As one can observe, all the three fields $\psi$, ${\hat a}_0$ and ${\hat a}_y$ are spatially modulated in the ${\hat x}$-direction. Among them, both $\psi$ and ${\hat a}_y$ have period $L$ as imposed by the boundary condition and they are $\frac{\pi}{2}$ out of phase with each other. The spatial modulation of the gauge field ${\hat a}_0$ is manifested, once we subtract the homogeneous part and consider $\triangle {\hat a}_0 = {\hat a}_0(\hat{x},u) - {\hat a}_0(0,u)$ as shown in figure \ref{fig:2b}. Unlike the other two fields, it turns out to have period equal to $L/2$. The solutions share the symmetry ${\hat x}\rightarrow -{\hat x}$ and ${\hat x}\rightarrow L/2 -{\hat x}$.

While solving the equations, the period $L$ can be chosen manually. However, it turns out that with $\mu$ and $b$ fixed, the free energy depends on the period $L$ and for a given $\mu$ there is a length $L$ for which the free energy shows a minimum. In order to see that, consider the system with fixed $\mu$ as the grand canonical ensemble and identify the free energy density with the Euclidean action evaluated on-shell. The free energy is given by
\begin{equation} \label{free-energy}
F = \frac{1}{N} S_{Euclidean}(\psi, a_0, a_y )|_{\text{on-shell}}.
\end{equation}
For each solution we have evaluated the free energy from (\ref{free-energy}), divide it by $L$ to obtain the free energy density and subtract the free energy density for the homogeneous solution from this.

\begin{figure}[h]
			\centering
			\begin{subfigure}[t]{7.5cm}
				\centering
				\includegraphics[width=7cm]{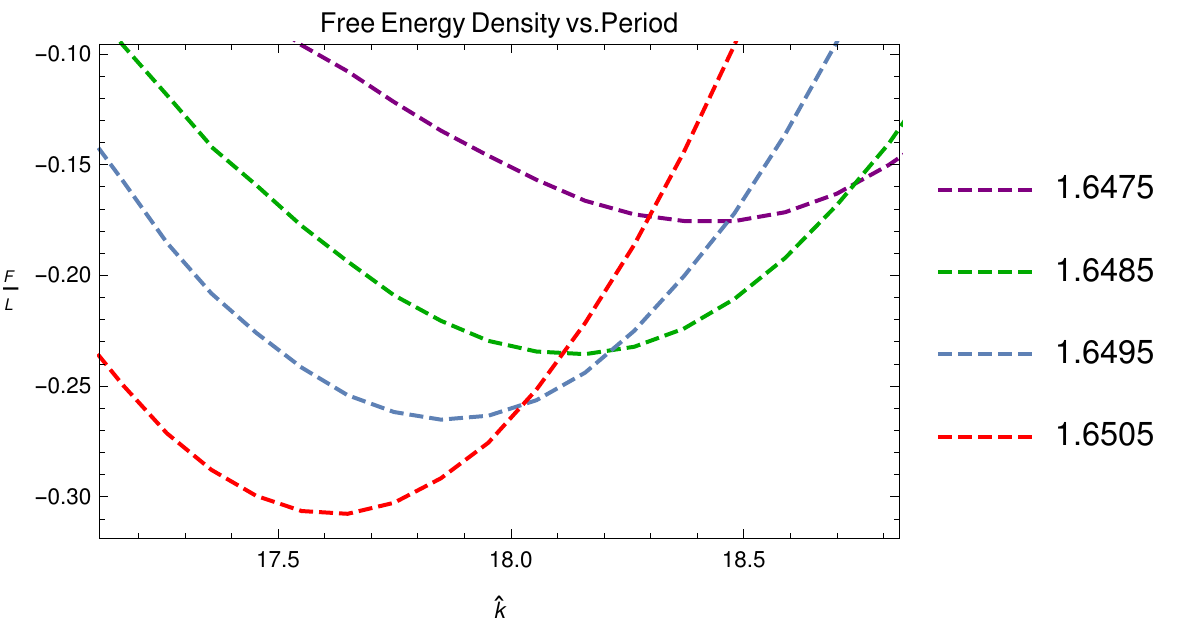}
						\caption{Free energy density $F/L$ vs. ${\hat k}$ for different values of the chemical potential $\mu$.}
				\label{fig:3a}
			\end{subfigure}%
			\quad
			\begin{subfigure}[t]{7.5cm}
				\centering
				\includegraphics[width=7cm]{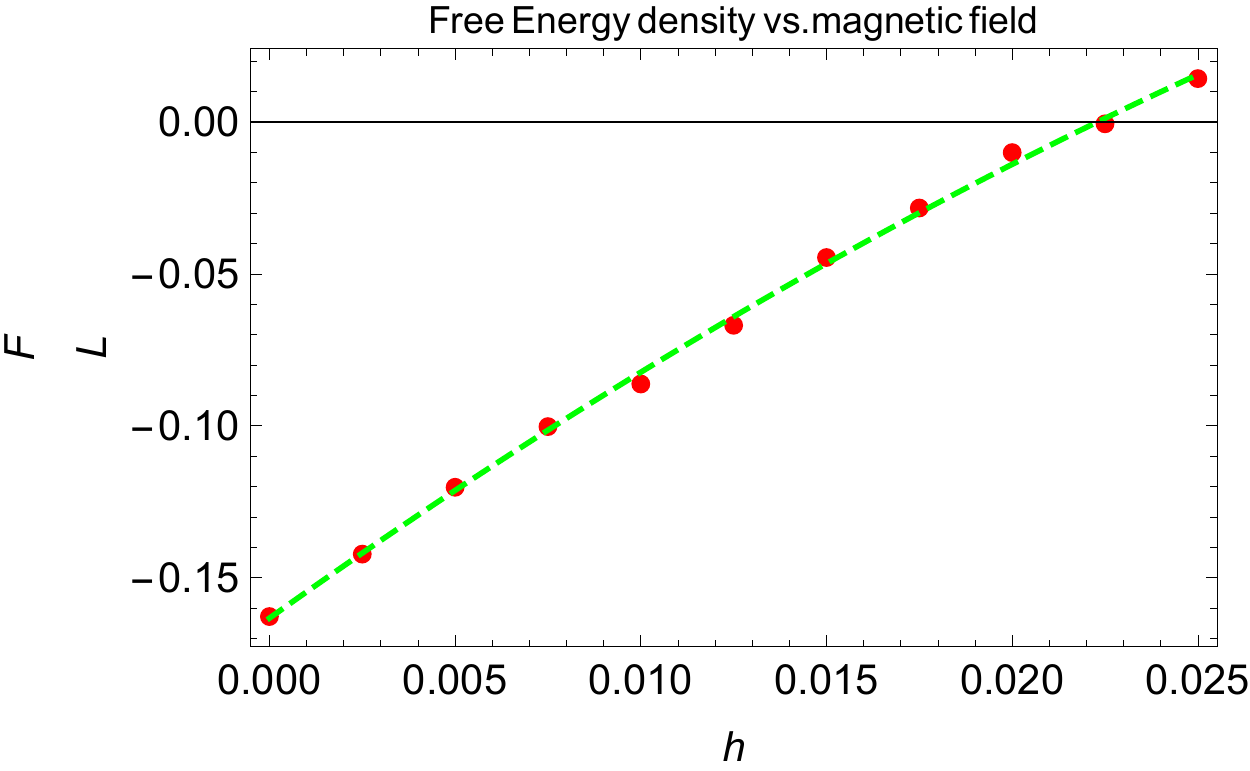}
						\caption{Free energy density $F/L$ vs. magnetic field $h$ for chemical potential fixed at $\mu=1.6475$ }
				\label{fig:3b}
			\end{subfigure}
			\caption{ }
			\label{fig:plot3}
		\end{figure}

We have varied the period $L$ in small increments and numerically solved the equations along with boundary conditions for a range of values of $L$.  We have plotted the free energy density $F/L$ vs ${\hat k}=\frac{2\pi}{L}$ for zero magnetic field in Fig.\ref{fig:3a} for several values of the chemical potential $\mu$.  Within the range of ${\hat k}$, it shows a minimum at around certain values of ${\hat k}={\hat k}_0$ depending on $\mu$ and for higher or lower values of ${\hat k}$ the free energy will increase. Since ${\hat k}={\hat k}_0$ corresponds to a minimum energy configuration and the corresponding free energy is less than that of the homogeneous solution, the spatially modulated configuration with wave number ${\hat k}={\hat k}_0$ is thermodynamically preferred. Forcing the system to have a higher or lower wave number will cost additional energy and will be thermodynamically unstable.

\begin{figure}[h]
			\centering
			\begin{subfigure}{8cm}
				\centering
				\includegraphics[width=7cm]{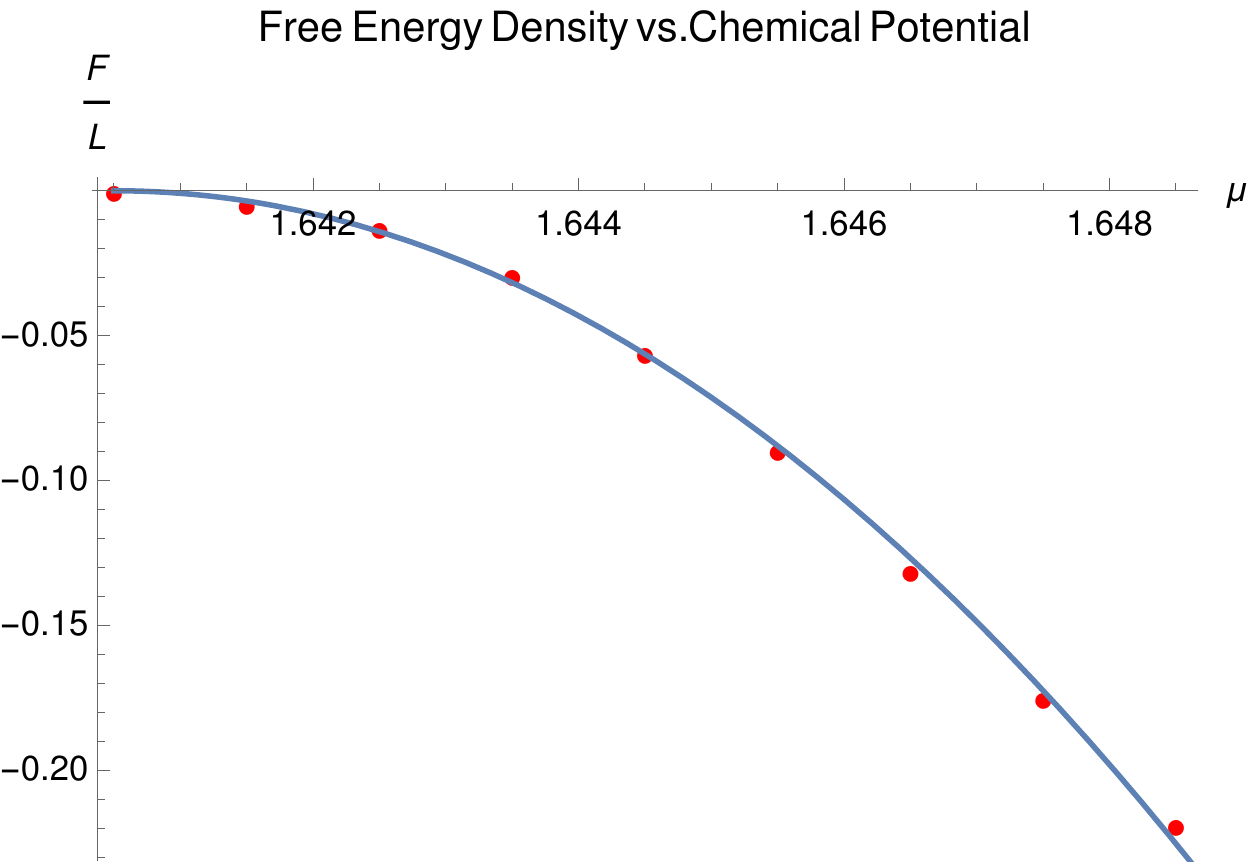}
						\caption{Free energy density $F/L$ vs. $\mu$ . \\The transition occurs at $\mu_c=1.6405$.}
				\label{fig:4a}
			\end{subfigure}%
			\begin{subfigure}{8cm}
				\centering
				\includegraphics[width=7cm]{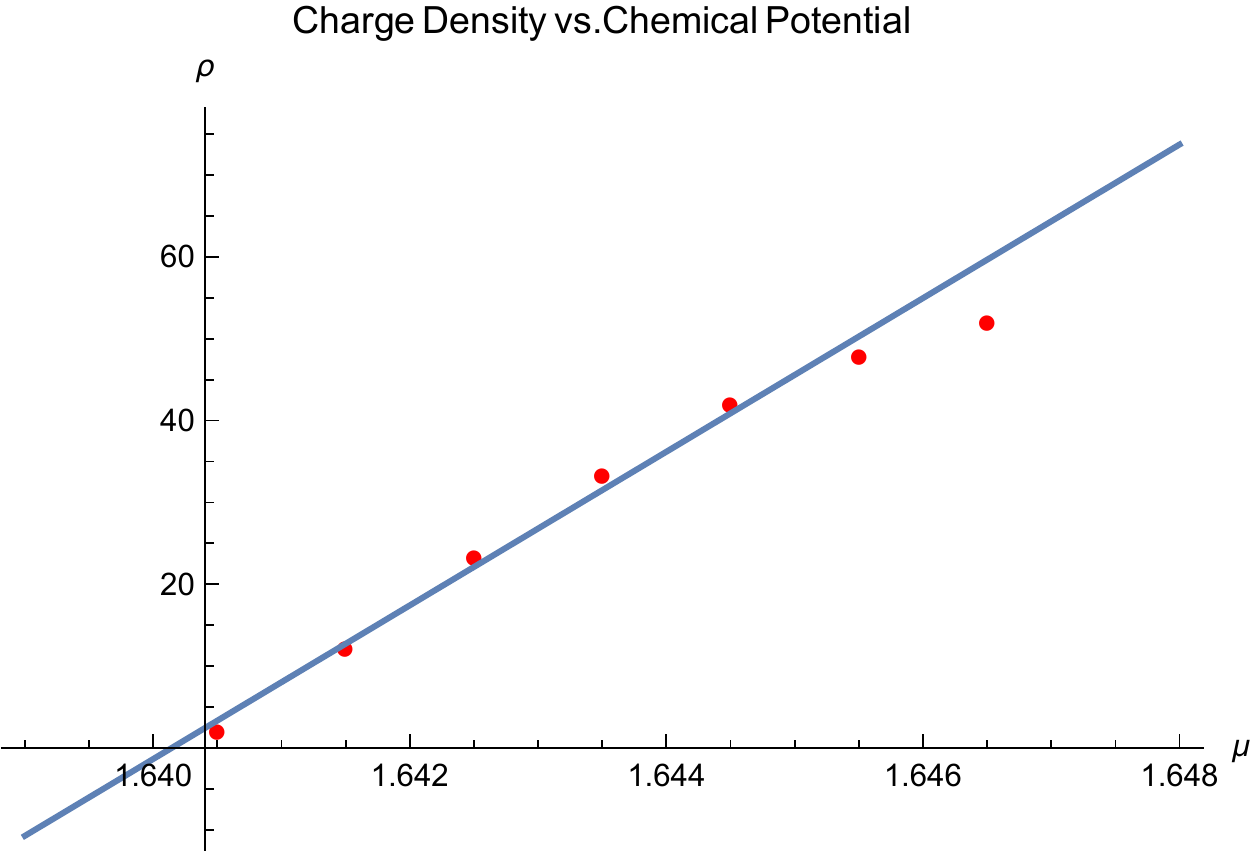}
						\caption{Charge density $\rho$ vs.  $\mu$.}
				\label{fig:4b}
			\end{subfigure}
			\caption{Plot of Free energy density $F/L$ and charge density $\rho$ vs.  $\mu$. }
			\label{fig:plot4}
		\end{figure}

Next, we have studied the behavior of the system with a variation of chemical potential. However, since there are multiple branches of solutions we wanted to focus on a single branch. For a specific branch of solution we expect that at the critical value of chemical potential both the amplitude of the charge density wave and the spin density wave, given by Max$(d(x) - <d>)$ and 
Max$(c_\psi - <c_\psi>)$ respectively should vanish and around that point they should have a smooth behavior.
%
For each $\mu$, we computed the free energy density for wave number ${\hat k}={\hat k}_0$, which corresponds to minimum free energy density.  This minimum free energy density (with the free energy density of the homogeneous solution subtracted) is plotted against $\mu$ in Fig.\ref{fig:4a}. As one can observe the free energy density increases with decreasing $\mu$ and at a critical value of the chemical potential $\mu=\mu_c$, which turns out to be $\mu_c=1.6405$ for the present choice of parameters, the free energy density of the inhomogeneous solution equals to that of the homogeneous solution indicating a phase transition. For $\mu > \mu_c$ the spatially modulated solution is thermodynamically preferred, while for $\mu < \mu_c$ the homogeneous solution is the stable solution. From the behavior of the plot, which turns out to have a quadratic behavior, $F/L \sim (\mu - \mu_c)^2$ we conclude that the nature of the phase transition is second order.  
The charge density for this system in the grand canonical ensemble is defined to be
 \begin{equation}\label{chargedensity}
 \rho = - \frac{\partial F}{\partial\mu}.
 \end{equation} 
This expression should be equal to the mean value of $d$ given in (\ref{asymbehave}), the expression of the charge density coming from the subleading term in the asymptotic expansion. From the plot of the free energy density vs. $\mu$ we obtain the plot of the charge density $\rho$ vs. $\mu$ in Fig.\ref{fig:4b} using (\ref{chargedensity}). The charge density at $\mu=\mu_c$ is non-zero and positive and is equal to that of the homogeneous configuration. Near $\mu=\mu_c$ the charge density increases linearly with $\mu$.

\begin{figure}[h]
			\centering
			\begin{subfigure}{8cm}
				\centering
				\includegraphics[width=7cm]{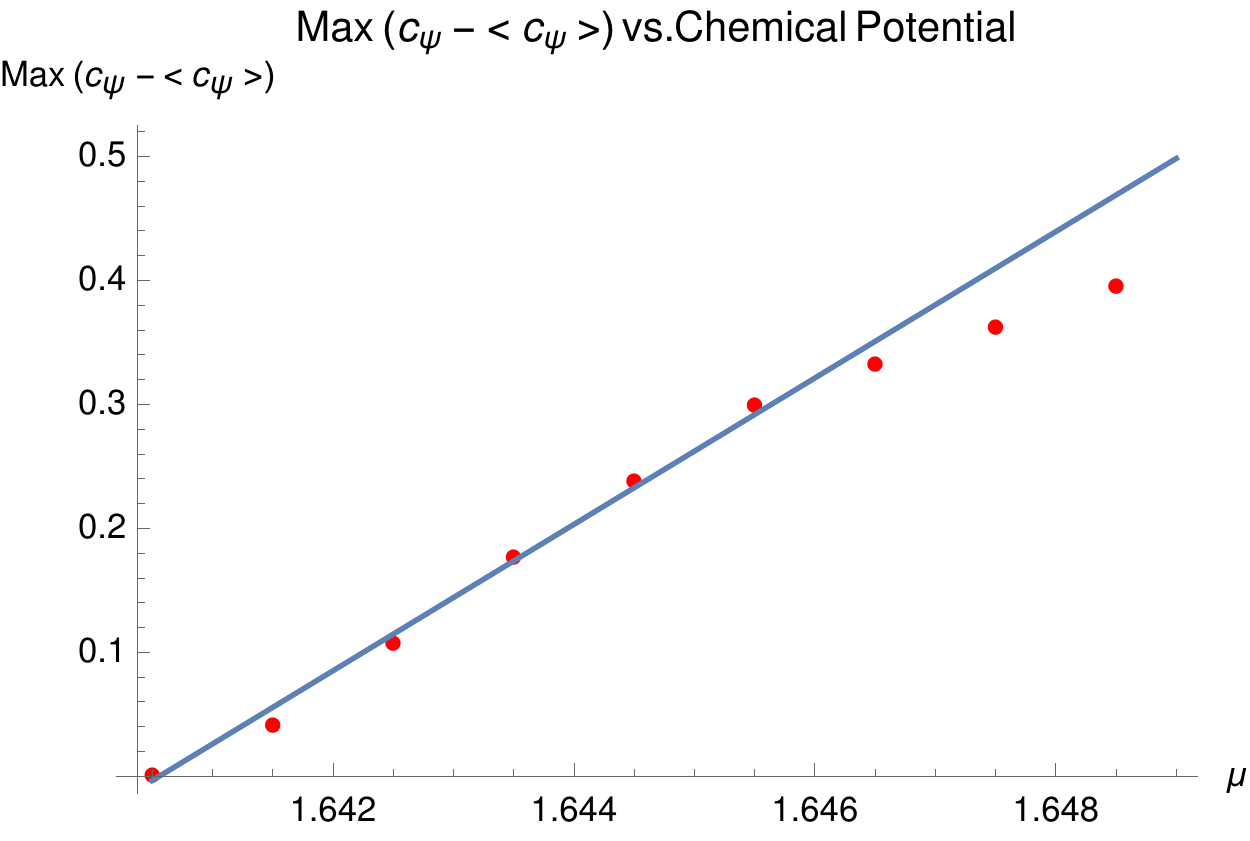}
						\caption{ Max($c_\psi-<c_\psi>)$ vs. $\mu$.}
				\label{fig:5a}
			\end{subfigure}%
		\begin{subfigure}{8cm}
				\centering
				\includegraphics[width=7cm]{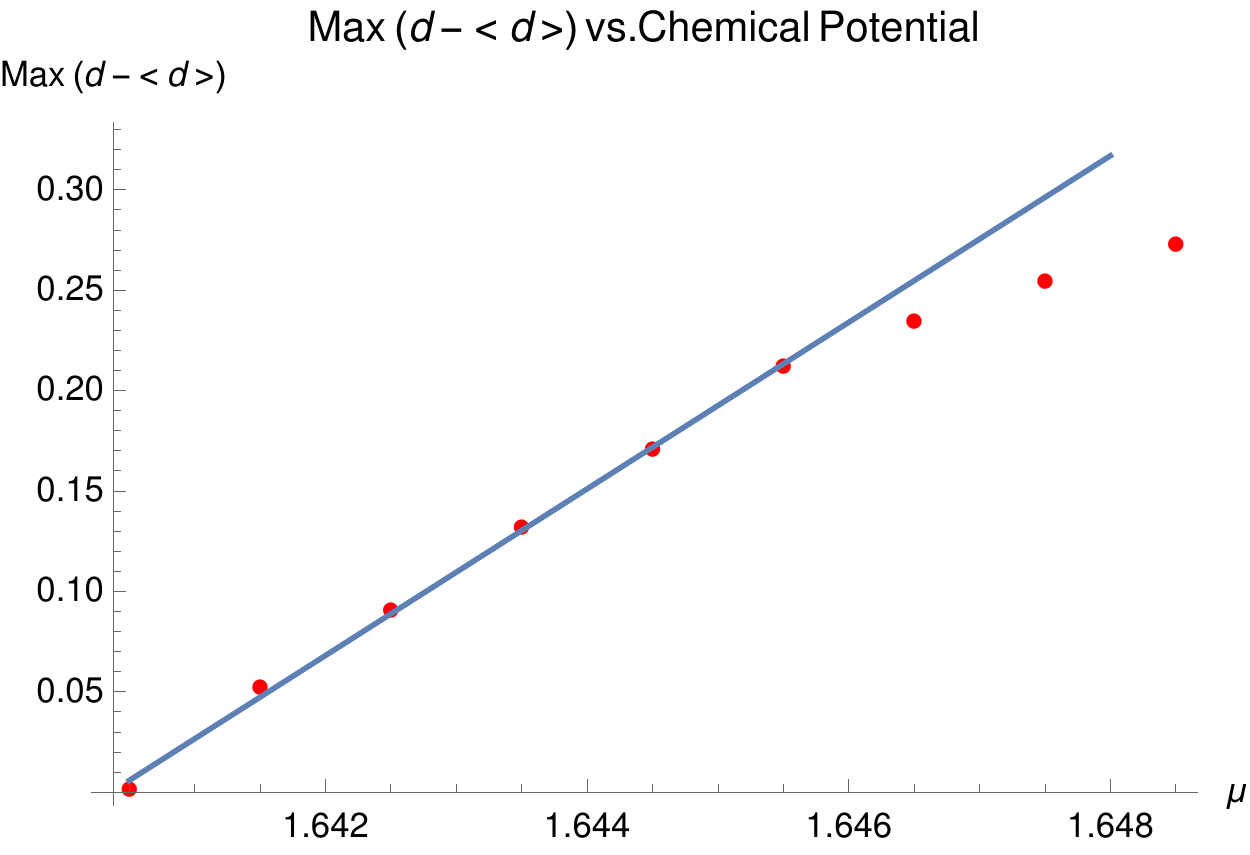}
						\caption{ Max$(d-<d>)$ vs. $\mu$.}
				\label{fig:5b}
			\end{subfigure}
			\caption{Amplitudes of the spin density and the charge density waves vs. $\mu$. }
			\label{fig:plot3}
\end{figure}

The amplitude of the charge density wave, given by Max$(d(x) - <d>)$ is plotted against the chemical potential $\mu$ in Fig. \ref{fig:5b}. Its behavior near $\mu_c$ is linear in $ (\mu - \mu_c)$.  
We have also plotted the Max$(c_\psi(x) - <c_\psi>)$ vs. $\mu$ in Fig.\ref{fig:5a}, which can be considered to be related to the amplitude of spin density wave \cite{10}. This also vanishes at $\mu=\mu_c$ and its behavior near $\mu_c$ is linear in $(\mu - \mu_c)$ unlike \cite{10}, where it was found to have a square root dependence. Comparing the plots of  Max$(d(x) - <d>)$ and Max$(c_\psi(x) - <c_\psi>)$ one can see amplitudes of the charge density wave and the spin density wave are of the same order in this case, unlike the result obtained in the case of D3-D7 model \cite{10}, where the amplitude of the spin density wave dominates. 
  
Next, we have turned on the magnetic field $h$ and find the spatially modulated solutions for different values of $L$, as in the case of $h=0$. In the presence of the magnetic field, however, ${\hat x}\rightarrow L/2 - {\hat x}$ symmetry is broken. We have considered the variation of the free energy density for the minimum energy configurations vs the magnetic field, which is plotted in Fig.\ref{fig:3b}. The plot shows the free energy density increases linearly with the magnetic field and (for a given $\mu$) there is a critical value of the magnetic field $h=h_c$ at which the free energy density of the spatially modulated configuration equals that of the homogeneous configuration. For $h>h_c$ the homogeneous solution has lower free energy density and so the spatially modulated configuration becomes thermodynamically unstable, while for $h<h_c$ it is the other way around. At the non-zero magnetic field the magnetization changes discretely with the transition indicating it is a first-order phase transition, unlike $h=0$.

\begin{figure}[h]
			\centering
			\begin{subfigure}{8cm}
				\centering
				\includegraphics[width=7cm]{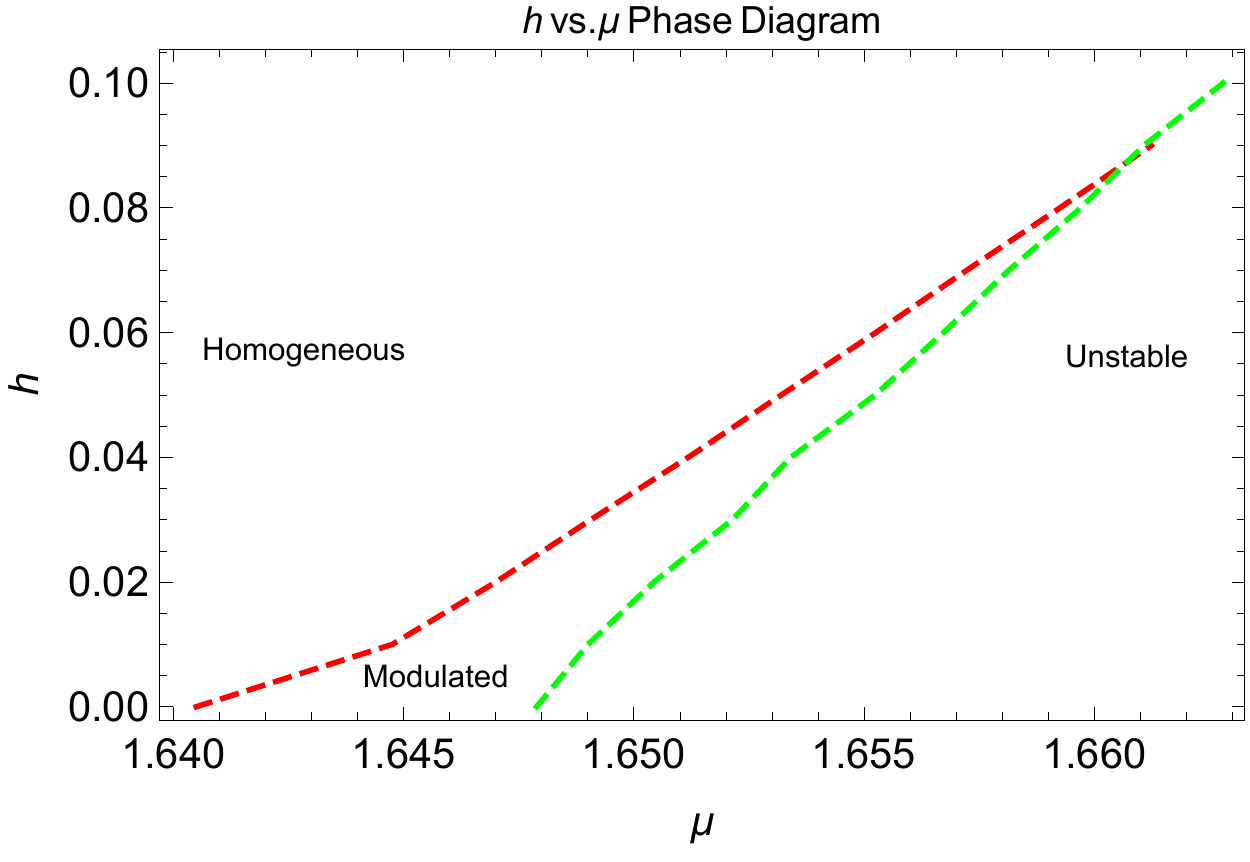}
						\caption{Magnetic field vs. Chemical potential.}
				\label{fig:6a}
			\end{subfigure}%
			\begin{subfigure}{8cm}
				\centering
				\includegraphics[width=7cm]{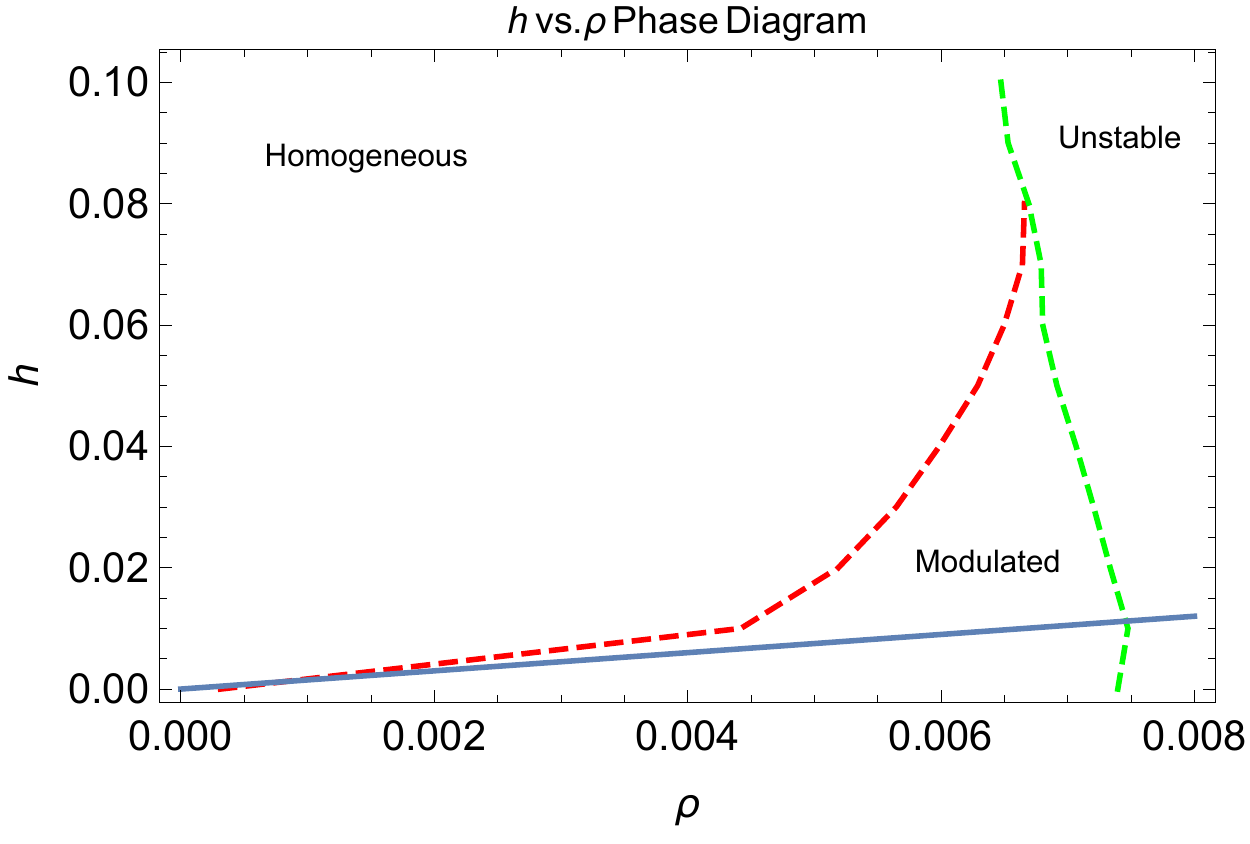}
						\caption{ Magnetic field vs. Charge density.}
				\label{fig:6b}
			\end{subfigure}
			\caption{Phase Diagrams}
			\label{fig:6}
\end{figure}
By varying both the chemical potential and the magnetic field, we have obtained the phase diagram in the $(\mu, h)$-plane, given in Fig.\ref{fig:6a}. For every value of the magnetic field $h$ we have computed the critical value of the chemical potential $\mu_c$, which we have plotted in the phase diagram. For a given value of the magnetic field, the homogeneous solution has lower free energy for smaller values of the chemical potential. As the chemical potential increases beyond a certain critical value, represented by the red dashed curve on the left of the Fig.\ref{fig:6a}, the modulated solution becomes thermodynamically favourable. However, as the chemical potential keeps on increasing, at a higher value $\frac{\partial \rho}{\partial \mu}$ vanishes and turns negative beyond that point, shown by the dashed green curve on the right in the Fig.\ref{fig:6a}, implying that the modulated solution becomes unstable to electrical fluctuation \cite{Chamblin:1999hg}. As the chemical potential increases further, it shows an unstable region corresponding to the right-hand side of the dashed green curve. It may be interesting to find out the configuration to which the electrical instability will lead to. If we keep on increasing the chemical potential even further, we find that once again the homogeneous solution becomes stable, in the sense that it will have lower free energy.

In order to compare the result with that of \cite{29}, we have presented a phase diagram in Fig.\ref{fig:6b} obtained by varying the charge density and the magnetic field in the $(\rho, h)$-plane. This also exhibits similar pattern as the phase diagram obtained by varying the magnetic field and the chemical potential.  For a given magnetic field, on the left part of the diagram, where charge density is small the homogeneous solution has lower free energy. As the charge density increases and crosses a critical value, represented by the red dashed curve on the left of the Fig.\ref{fig:6b}, the modulated solution starts having lower free energy. With further increase of the charge density, however, it hits the dashed green curve, from where  $\frac{\partial \rho}{\partial \mu}$ turns negative signalling begining of the region where the modulated solution is electrically unstable. As mentioned in \cite{29}, the quantum Hall state is a stable configuration, which occurs at $h=\frac{3 \rho}{2}$, which we have represented in this figure by the solid blue line. We have chosen an overall normalisation factor for the charge density $\rho$ so that the region of stablity occurs on the top left of the solid blue line. Comparing with  the result of that of \cite{29}, we observe that since the value of the mass parameter in our case is smaller, $m=0.5$ compared to $m=1$ in their case, the stability region of the modulated solution is much closer to the solid blue line, as mentioned there. For zero magnetic field, the red dashed curve starts from non-zero charge density at the critical point (which is not clear in the figure). There is a critical value of the charge density where it enters the region of electrical instability. We find that this region of electrical instability is also present for the region $h<3 \rho/2$, which is different from the result of \cite{29}.
 
\section{Discussion}

In this work, we have considered a probe D2-D8 brane model and solving non-linear partial differential equations following from the action along with the boundary conditions, we have shown that the equations admit spatially modulated solution. In order to avoid the conical singularity, we have chosen the mass parameter $m$ to be non-zero. Plotting the free energy density vs. the wave number ${\hat k}=\frac{2\pi}{L}$ we find, within the range of parameters, it shows a minimum at certain ${\hat k}={\hat k}_0$, indicating spontaneous breakdown of translational symmetry. This spatially modulated solution is thermodynamically preferred for the value of the chemical potential $\mu > \mu_c$.  From the behavior of the amplitude of the charge density and the $c_\psi$, it turns out that this spatially modulated solution involves both a spin density wave and a charge density wave. The behavior of the free energy density with the variation of the chemical potential, in the absence of the magnetic field, indicates a second order phase transition occurring at $\mu=\mu_c$. While studying the behavior of the free energy density with the variation of the magnetic field, we find that the free energy density increases linearly with the magnetic field and for a given value of the chemical potential, there is a critical value of the magnetic field where the spatially modulated solution makes a first-order phase transition to the homogeneous configuration. The phase diagram obtained by variation of the chemical potential and the magnetic field shows that for a given value of the magnetic field, there is a critical value of the chemical potential below which the homogeneous configuration is preferred. As the chemical potential increases, it encounters an electrical instability at a higher value as well. So for large magnetic field or large chemical potential, the spatially modulated configuration is not stable.

Similar spatially modulated solutions have been obtained for Einstein-Maxwell-dilaton or axion systems in the bottom-up approach \cite{13,14,15}, which are dual to charge/current density wave and for the D3-D7 model in the top-down approach \cite{10}, which is a mixture of a spin density wave and a charge density wave. In that respect, this solution is more similar to that obtained in the D3-D7 model. However, our results differ from that of the D3-D7 model on a few scores. Unlike the D3-D7 model, in the present case the amplitudes of both the charge density and spin density waves are comparable while spin density wave dominates in D3-D7 model. In the D2-D8 model, near the critical value of the chemical potential, amplitude of the spin density wave exhibits a behavior linear in $\mu-\mu_c$, where in the D3-D7 model it was found to be $\sqrt{\mu-\mu_c}$. Considering the phase diagram in the $(\mu , h)$ plane in Fig.\ref{fig:6a} we find that for a given value of the magnetic field, as the chemical potential increases, the inhomogenous solution encounters electrical instability, in contrast with the D3-D7 model. In particular, in the D2-D8 model the charge and spin density wave is unstable for large values of the chemical potential and the magnetic field.  

This work can be further extended in several directions. Study of the direct conductivity as well as the thermoelectric properties along the line of \cite{11} and their dependence on temperature and other parameters may enable one to connect it to the experimental observations and provides a better theoretical understanding. It would also be interesting to consider the dual field theory and understand the source of the electrical instability.

Another direction is to study the fermionic responses of this system along the line of \cite{DeWolfe:2011aa,DeWolfe:2012uv,DeWolfe:2014ifa,self,Cosnier-Horeau:2014qya,self1}. Some works have appeared for striped solution obtained in Einstein-Maxwell-dilaton model with two $U(1)$ gauge fields \cite{17} in a bottom-up approach. Solving the Dirac equation in the background with the present solution one can study the Fermi surfaces, the conditions when the Fermi surface will form or spectrum will be gapped and the roles of the parameters determining its properties.
 
 One can also investigate the role of symmetry breaking mechanisms in the present context. For spontaneous symmetry breaking it gives rise to Goldstone modes, which may lead to sliding of the stripes under a weak electric field. In addition, if we incorporate explicit symmetry breaking, the Goldstone modes are lifted and the stripes remain pinned. However, for large enough electric fields such solutions may lead to a non-linear behavior of conductivity \cite{11,38}, which can be studied in the present context.
Finally, in the present work, we have used probe approximation and neglected the back reaction of the gravity. One can incorporate the back reaction by considering the solution of Einstein equation as well to obtain a spatially modulated solution in the present set up. We hope to report about some of these issues in the near future. 
%
			\section*{Acknowledgement}
We thank the anonymous referee for his/her useful suggestions and comments. SM thankfully acknowledges the assistance received from Science and Engineering Research Board (SERB), India (project file no. CRG/2019/002167).
\newpage

\bibliographystyle{unsrt}

		\end{document}